\pdfoutput=1
\documentclass[prb,twocolumn,showpacs,amsmath,amssymb,floatfix,superscriptaddress]{revtex4}

\usepackage{graphicx}
\usepackage{dcolumn}
\usepackage{bm}

\begin{document}

\title{Spin-Seebeck effect on the surface of a topological insulator due to nonequilibrium spin-polarization parallel to the direction of thermally driven electronic transport}

\author{Po-Hao Chang}
\affiliation{Department of Physics and Astronomy, University of Delaware, Newark, DE 19716-2570, USA}
\author{Farzad Mahfouzi}
\affiliation{Department of Physics and Astronomy, University of Delaware, Newark, DE 19716-2570, USA}
\author{Naoto Nagaosa}
\affiliation{RIKEN Center for Emergent Matter Science (CEMS), Wako, Saitama 351-0198, Japan}
\affiliation{Department of Applied Physics, University of Tokyo, Tokyo 113-8656, Japan}
\author{Branislav K. Nikoli\' c}
\email{bnikolic@udel.edu}
\affiliation{Department of Physics and Astronomy, University of Delaware, Newark, DE 19716-2570, USA}
\affiliation{RIKEN Center for Emergent Matter Science (CEMS), Wako, Saitama 351-0198, Japan}

\begin{abstract}

We study the transverse spin-Seebeck effect (SSE) on the surface of a three-dimensional topological insulator
(TI) thin ﬁlm, such as Bi$_2$Se$_3$ , which is sandwiched between two normal metal leads. The temperature bias
$\Delta T$ applied between the leads generates surface charge current which becomes spin polarized due to strong
spin-orbit coupling on the TI surface, with polarization vector acquiring a component $P_x \simeq 60\%$ parallel to the
direction of transport. When the third nonmagnetic voltage probe is attached to the portion of the TI surface
across its width $L_y$, pure spin current will be injected into the probe where the inverse spin Hall effect (ISHE)
converts it into a voltage signal \mbox{$|V_\mathrm{ISHE}|^\mathrm{max}/\Delta T \simeq 2.5$ $\mu$V/K} (assuming the SH angle of the Pt voltage probe and
$L_y=1$ mm). The existence of predicted nonequilibrium spin polarization parallel to the direction of electronic
transport and the corresponding electron-driven SSE crucially relies on orienting quintuple layers (QLs) of Bi$_2$Se$_3$ orthogonal to the TI surface and tilted by $45^\circ$ with respect to the direction of transport. Our analysis is based on
the Landauer-B\"{u}ttiker-type formula for spin currents in the leads of a multiterminal quantum-coherent junction, which is constructed by using nonequilibrium Green function formalism within which we show how to take into account arbitrary orientation of QLs via the self-energy describing coupling between semi-inﬁnite normal metal leads and the TI sample.

\end{abstract}

\maketitle

\section{Introduction}\label{sec:introduction}

The spin-Seebeck effect (SSE) is a recently observed phe-
nomenon where spin current or spin accumulation is induced
by a temperature gradient applied across a ferromagnetic
material.~\cite{Bauer2012,Uchida2012}. At ﬁrst sight, the SSE appears to be a counterpart of the traditional charge-Seebeck effect (CSE) where the
temperature gradient across a conductor induces electrical
current (in closed circuits) or voltage (in open circuits).
However, the SSE has been observed in a surprisingly wide
range of materials, including ferromagnetic insulators where
the CSE does not exist. ~\cite{Bauer2012,Uchida2012}


%
\begin{figure}
\includegraphics[scale=0.12,angle=0]{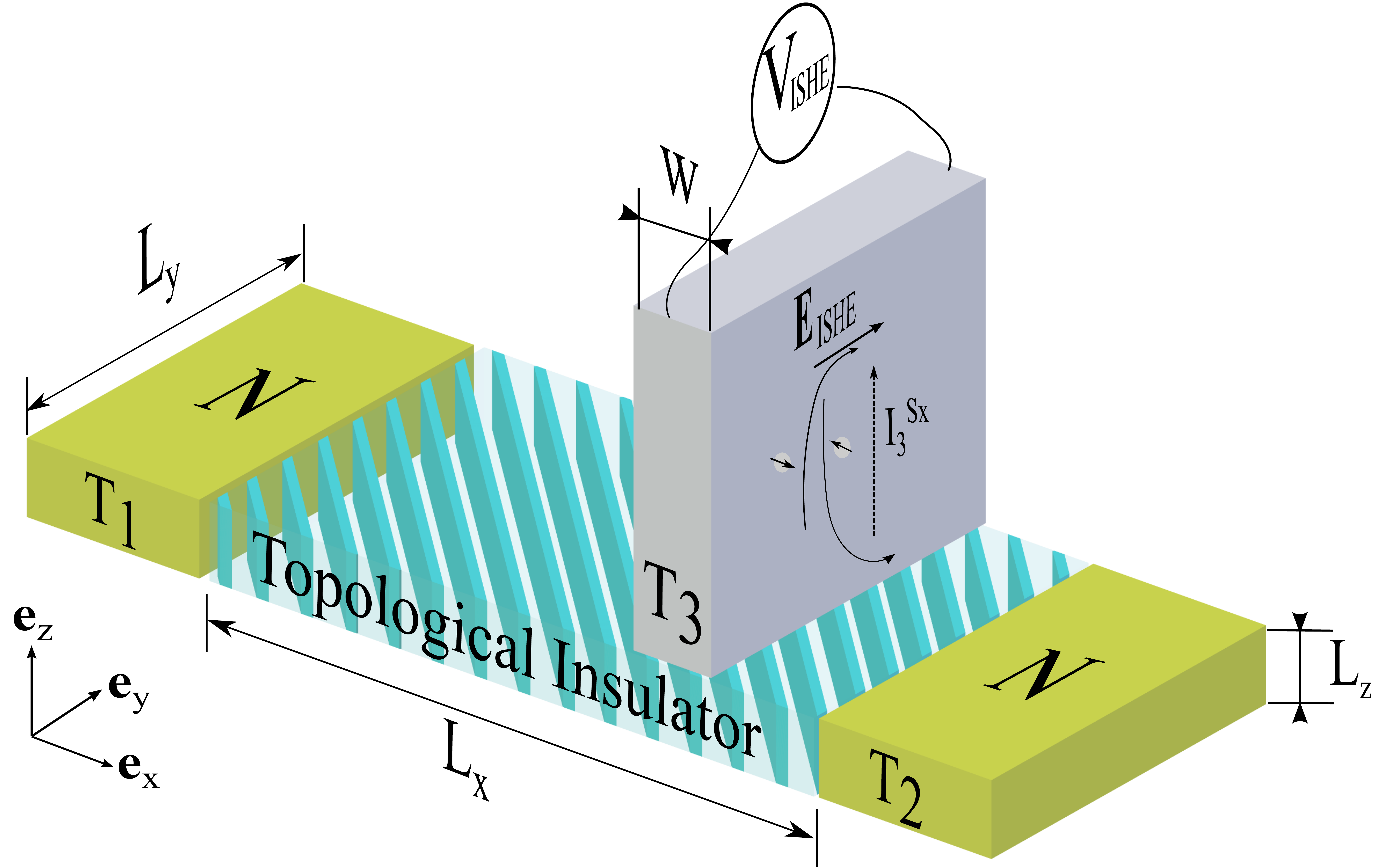}
\caption{(Color online) Schematic view of the three-terminal junc-
tion where a thin ﬁlm of Bi$_2$Se$_3$ is attached to two nonmagnetic
metallic leads kept at different temperatures $T_1>T_2$. The SSE voltage
signal $V_\mathrm{ISHE}$ is measured across the edges of the third nonmagnetic
metallic lead attached to the top TI surface a function of its position
along the x axis. This lead acts as voltage probe attached to a
macroscopic reservoir at potential $V_3$ to ensure zero net charge current
through it. We orient quintuple layers of Bi$_2$Se$_3$ at an angle of $45^\circ$
with respect to the direction ($x$ axis) of electron propagation on the
TI surface that is orthogonal to these layers.}
\label{fig:fig1}
\end{figure}

In the so-called transverse SSE measurement geometry,
illustrated in Fig.~\ref{fig:fig1}, a temperature gradient applied longitudinally over a strip of magnetic material induces a spin signal
detected by measuring voltage $V_\mathrm{ISHE}$ generated via the inverse
spin Hall effect (ISHE) in the nonmagnetic metallic probe
(such as Pt) attached on top of the strip across its width. The
voltage signal $V_\mathrm{ISHE}$ is found to be approximately linear (in
ferromagnetic metals and insulators~\cite{Uchida2012}) or a hyperbolic sine
(in ferromagnetic semiconductors~\cite{Jaworski2010}) function of the probe
position in the longitudinal direction over the length of several
millimeters. Since this is surprisingly long when compared
to the usual electronic spin-dependent length scales, recent
theories of the SSE have focused on the interplay of magnons
and phonons out of equilibrium.~\cite{Adachi2013}.

On the other hand, {\em the role of spin-polarized conduction electrons in SSE generation has been much less explored}.
The need for this has been prompted by the very recent
experimental~\cite{Jaworski2012} unveiling of ``giant'' $V_\mathrm{ISHE}$ (up to a thousand
times larger than observed in measurements on magnetic
materials~\cite{Uchida2012}) in the transverse SSE setup where a {\em nonmagnetic}
semiconductor InSb was placed in a large longitudinal (parallel
to the temperature gradient) external magnetic ﬁeld. Aside
from the magnitude of $V_\mathrm{ISHE}$, which is speculated to arise in the
interplay of spin-orbit coupling (SOC) and enhancement of the
phonon drag contribution to both the spin- and charge-Seebeck
coefﬁcients for electrons pushed into the ultraquantum limit
by the applied magnetic ﬁeld, another puzzle for SSE theories
posed by Ref.~\onlinecite{Jaworski2012} is that $V_\mathrm{ISHE}$ did not change sign under the reversal of the magnetic-ﬁeld direction.

Here we show that basic phenomenology of the experiment in Ref.~\onlinecite{Jaworski2012} can be recreated without applying any external
magnetic ﬁeld. The role of the magnetic ﬁeld was to spin
polarize electrons in the direction of transport (by means of
Zeeman splitting, further ampliﬁed by SOC in InSb), as well
as to conﬁne their spatial motion so that electrons spiral in
the $yz$ plane (with cyclotron orbits that are quantized into
Landau levels) as they translate along the $x$ axis. In order to
generate the same spin polarization along the electron transport
direction, we employ a thin ﬁlm of a recently discovered three-
dimensional topological insulator (3D TI) material, such as
Bi$_2$Se$_3$ assumed here, which is attached to three normal (i.e.,
nonmagnetic) metal (N) leads, as illustrated in Fig.~\ref{fig:fig1}.

The 3D TI materials~\cite{Hasan2010} possess a usual band gap in the bulk, while hosting metallic surfaces whose low-energy quasiparticles are massless Dirac fermions with spins locked to their momenta due to strong Rashba-type SOC.~\cite{Winkler2003} In particular, Bi$_2$Se$_3$ realization of TI is a strongly anisotropic material composed of  quintuple layers (QLs) of Bi and Se atoms, where one QL consists of three Se layers strongly bonded to two Bi layers in between.~\cite{Hasan2010} While Bi$_2$Se$_3$ is always unintentionally $n$-type doped by Se vacancies, charge carriers in the bulk of films of thickness $\lesssim 10$ nm can be completely removed by a gate electrode.~\cite{Kim2012}

One of the key effects~\cite{Burkov2010,Misawa2011,Pesin2012,Modak2012} that 3D TIs bring into spintronics is nonequilibrium spin density in the direction {\em transverse} to injected unpolarized charge current, which is much larger~\cite{Pesin2012} than in the case~\cite{Edelstein1990,Inoue2003,Ganichev2006a} of two-dimensional electron gases (2DEGs) with the Rashba SOC. Our first principal result [see Figs.~\ref{fig:fig2}(a) and ~\ref{fig:fig2}(b)] demonstrates that additional component of nonequilibrium spin density and polarization can be induced in the direction {\em parallel} to injected charge current, on the proviso that QLs are oriented as shown in Fig.~\ref{fig:fig1}. Our second principal result [see Figs.~\ref{fig:fig4}(a)--(c)] shows that this indeed makes possible non-zero SSE signal in three-terminal geometry of the junction depicted in Fig.~\ref{fig:fig1}.

The paper is organized as follows: In Sec.~\ref{sec:lbformulas} we employ
the nonequilibrium Green function (NEGF) formalism~\cite{Stefanucci2013} to obtain the Landauer-B\"{u}ttiker (LB-type) formula for spin currents in the leads of a multi-terminal quantum-coherent junction driven by both voltage bias and temperature bias in the linear-response regime. Section~\ref{sec:hamiltonian} explains our Hamiltonian model for the TI ﬁlm, as well as the construction of the retarded GF for an open system TI + semi-inﬁnite N leads where the orientation of QLs shown in Fig.~\ref{fig:fig1} is taken into account through the self-energy entering the retarded GF. In Sec.~\ref{sec:polarization} we analyze
the spin-polarization vector of the charge current, as well as charge conductance, for a TI ﬁlm attached to two N leads. In Sec.~\ref{sec:sse} we predict the magnitude of the voltage signal generated across the third N lead in the three-terminal junction depicted
in Fig. 1, while also contrasting its features with those of conventional charge and spin-dependent Seebeck coefﬁcients that would be measured between terminals 1 and 2 in Fig.~\ref{fig:fig1}. We conclude in Sec.~\ref{sec:conclusions}.

\section{Spin currents in multi-terminal quantum-coherent conductors driven by voltage and temperature biases}\label{sec:lbformulas}

We use the same units for the total charge $I_p=I^\uparrow_p + I^\downarrow_p$ and total spin $I_p^{S_\alpha} = I^\uparrow_p - I^\downarrow_p$ currents ﬂowing through lead $p$,
which are constructed from spin-resolved charge currents $I_p^\sigma$
with the spin quantization axis for $\sigma=\uparrow,\downarrow$ chosen along $\mathbf{e}_\alpha$. There has been a lively debate~\cite{Scheid2007,Nikolic2005} in the literature on the proper derivation of the multi-terminal LB-type formula~\cite{Datta1995} which connects spin current $I_p^{S_\alpha}$ ﬂowing through the semi-inﬁnite ideal (i.e., charge- and spin-interaction-free) metallic lead p attached to a quantum-coherent conductor due to voltages $V_p$ applied at the external macroscopic reservoirs into which the leads terminate at inﬁnity. The debate was spurred by one of the early derivations,~\cite{Pareek2004} using the traditional scattering
matrix framework [19], which predicted unphysical $I_p^{S_\alpha} \neq 0$
in equilibrium $V_p = \mathrm{const}$.


Here we derive LB-type formula for spin currents driven by
{\em both} voltage and temperature biases. We bypass the issue of
unphysical equilibrium total spin currents~\cite{Scheid2007,Nikolic2005} by starting
from the outset from a general NEGF-based expression for
spin current in lead $p$:
\begin{equation} \label{eq:mw}
I_p^{S_\alpha} =  \frac{e}{h} \int \!\! dE \, \mathrm{Tr} \, \{ \hat{\sigma}_\alpha [{\bm \Sigma}_p^<(E) {\bf G}^>(E) - {\bm \Sigma}^>_p(E) {\bf G}^<(E)] \}.
\end{equation}
This is actually the difference of spin-resolved charge currents
given by the well-known Meir–Wingreen formula,~\cite{Stefanucci2013} where $(\hat{\sigma}_x,\hat{\sigma}_y,\hat{\sigma}_z)$ is the vector of the Pauli matrices. The two fundamental objects of the NEGF formalism—the retarded $\mathbf{G}(E)$
and the lesser \mbox{$\mathbf{G}^<(E)=\mathbf{G}(E) {\bm \Sigma}^<(E) \mathbf{G}^\dagger(E)$} GFs---describe the density of available quantum states and how electrons
occupy those states, respectively.~\cite{Stefanucci2013}

In the elastic transport regime the lesser self-energies,\mbox{${\bm \Sigma}^<_p(E) = i f_p(E) {\bm \Gamma}_p(E)$} and  ${\bm \Sigma}^< = \sum_p {\bm \Sigma}^<_p(E)$, are express-
ible in terms of the retarded self-energies ${\bm \Sigma}_p(E)$ using\mbox{${\bm \Gamma}_p=i [{\bm \Sigma}_p(E) - {\bm \Sigma}^\dagger_p(E)]$} and $f_p(E)$ as the Fermi distribution of electrons within the reservoirs. This makes it possible to
rewrite Eq.~\eqref{eq:mw} for the total spin current in lead $p$ as
\begin{widetext}
\begin{eqnarray} \label{eq:lbspin}
I_{p}^{S_\alpha}  =   \frac{e}{h} \sum_{q} \int \!\! dE \, \mathrm{Tr} \, [\hat{\sigma}_{\alpha} {\bm \Gamma}_{q}(E)\mathbf{G}(E)\mathbf{\bm \Gamma}_{p}(E)\mathbf{G}^{\dagger}(E)] \left\{ f_{p}(E)-f_{q}(E) \right\}.
\end{eqnarray}
By expanding \mbox{$f_p(E)-f_q(E)$} to linear order in \mbox{$T_{p}-T_{q}$} and
\mbox{$V_{p}-V_{q}$}, we ﬁnally get the desired multiterminal LB-type
formula for spin current driven by both temperature and
voltage biases in the linear-response regime:
\begin{eqnarray} \label{eq:lbtempspin}
I_{p}^{S_\alpha}  = \frac{e^2}{h} \sum_{q}\int \!\! dE \,  \mathrm{Tr}\, [\hat{\sigma}_\alpha\mathbf{\Gamma}_{q}(E)\mathbf{G}(E)\mathbf{\Gamma}_{p}(E)\mathbf{G}^{\dagger}(E)]  \left\{ \frac{\partial f}{\partial E} \left[ \frac{E-E_F}{eT}(T_p - T_q) - (V_p - V_q) \right] \right\}.
\end{eqnarray}
Note that the usual expression~\cite{Datta1995} for the total charge current
$I_p$ in lead $p$ is the same as Eq.~\eqref{eq:lbtempspin}, except that $\hat{\sigma}_{\alpha} \mapsto \hat{\sigma}_0$ where
 $\hat{\sigma}_0$ is the unit $2 \times 2$ matrix.

Applying Eq. (3) to the three-terminal junction in Fig..~\ref{fig:fig1}
gives
\begin{eqnarray} \label{eq:lbsee}
I_{p}^{S_\alpha}  = \frac{e^2 L_y}{2\pi h} \sum_{q=1}^3 \int \!\! \int \!\! dE\, dk_y\, \mathrm{Tr}\, [\hat{\sigma}_\alpha\mathbf{\Gamma}_{q}(E)\mathbf{G}(E)\mathbf{\Gamma}_{p}(E)\mathbf{G}^{\dagger}(E)]  \left\{ \frac{\partial f}{\partial E} \left[ \frac{E-E_F}{eT}(T_p - T_q) - (V_p - V_q) \right] \right\},
\end{eqnarray}
\end{widetext}
where we use $T_1 > T_2$, $T_3 (x)=T_1 - x(T_1-T_2)/L_x$, and
$V_1=V_2 \neq V_3$. By imposing the condition $I_p=0$ in one of
the leads, such as lead $p=3$ in Fig.~\ref{fig:fig1}, the linear system of
equations in Eq. (4) can be solved to ﬁnd voltage $V_3$ that has to
be applied to convert this lead into a voltage probe employed
in SSE experiments.



Using the spin-dependent transmission function \mbox{$\mathcal{T}_{21}^\alpha(E) = \mathrm{Tr}\, [\hat{\sigma}_\alpha {\bm \Gamma}_{2}(E)\mathbf{G}(E)\mathbf{\bm \Gamma}_{1}(E)\mathbf{G}^{\dagger}(E)]$} of the two-terminal version of junction in Fig.~\ref{fig:fig1}, we can compute the following integrals~\cite{Sivan1986,Nikolic2012}
\begin{equation}\label{eq:kintegral}
K_n^\alpha(\mu) = \frac{1}{h} \int\limits_{-\infty}^{\infty} dE\, \mathcal{T}_{21}^\alpha(E)  (E - E_F)^n \left(-\frac{\partial f}{\partial E} \right),
\end{equation}
which yield the three spin-dependent Seebeck coefficients \mbox{$S_\mathrm{spin}^\alpha=K_1^\alpha/(eTK_0^\alpha)$} for the chosen Pauli matrix $\hat{\sigma}_\alpha$, or the charge-Seebeck coefficient \mbox{$S_\mathrm{charge}=K_1/(eTK_0)$} when \mbox{$\hat{\sigma}_\alpha \mapsto \hat{\sigma}_0$}.

\section{Hamiltonian and retarded Green function for multi-terminal TI-based junctions}\label{sec:hamiltonian}

The junction in Fig.~\ref{fig:fig1} is modeled on the simple cubic lattice with lattice spacing $a$, which is assumed to be periodically repeated in the $y$-direction. The TI thin film has finite length $L_x$, while it is sufficiently thick $L_z=30a$ to ensure no coupling between the top and the bottom metallic surfaces which penetrate as evanescent states into the bulk of the TI film and whose overlap would open a minigap at the Dirac point (DP) in ultrathin films.~\cite{Park2010} The TI thin film is described using the minimal tight-binding Hamiltonian with four orbitals per site~\cite{Liu2010}
\begin{eqnarray}\label{eq:ti}
\mathbf{H}_\mathrm{TI} & = & \sum_{n,k_y} (\mathbf{c}_{n,k_y}^{\dagger} [\mathbf{M}_{0} + C\mathbf{1} + \mathbf{T}_{y}e^{ik_{y}a} + \mathbf{T}_{y}^{\dagger} e^{-ik_{y}a}] \mathbf{c}_{n,k_y}) \nonumber \\
\mbox{} && + \sum_{n, k_y, \alpha=x,z} (\mathbf{c}_{n,k_y}^{\dagger} \mathbf{T}_{\alpha} \mathbf{c}_{n + \mathbf{e}_\alpha,k_y} + \mathrm{H.c.}),
\end{eqnarray}
where \mbox{$\mathbf{T}_{\alpha} = B \hat{\sigma}_{z} \otimes \sigma_{0} - i A \hat{\sigma}_{x} \otimes \hat{\sigma}_{\alpha}/2$}; $\mathbf{M}_{0}=(M-6B) \hat{\sigma}_{z} \otimes \hat{\sigma}_{0}$; $\hat{\sigma}_0$ is the unit $2 \times 2$  matrix; and $\mathbf{1}=\hat{\sigma}_0 \otimes \hat{\sigma}_0$. Here ${\bf c}_{n,k_y}=(\hat{c}_{+\uparrow}, \hat{c}_{+\downarrow}, \hat{c}_{-\uparrow}, \hat{c}_{-\downarrow})^T$ annihilates electron in different orbitals on site $n$ with the transverse momentum $k_y$. The numerical values of the parameters are chosen as: \mbox{$M=0.3$ eV}; \mbox{$A=0.5$ eV}; and \mbox{$B=0.25$ eV}. The bottom of the band of TI is shifted by \mbox{$C=3.0$ eV}.

When applying the in-plane external magnetic field $B_x$, the Zeeman term changes to ${\bf M}_{0} \mapsto  {\bf M}_{0} + \Delta \hat{\sigma}_{0} \otimes \hat{\sigma}_{x}$ where $\Delta = g_\parallel \mu_{B} B_x$. We also apply an additional Zeeman term $\Delta \hat{\sigma}_{0} \otimes \hat{\sigma}_{z}$ with $\Delta =0.5$ eV to the bottom TI surface in order to split its Dirac cone and block current through it.

The semi-infinite N leads made of nonmagnetic metallic material are described by a tight-binding Hamiltonian with a single orbital per site
\begin{eqnarray} \label{eq:n}
\mathbf{H}_N & = & \sum_{n, \sigma,k_y} \varepsilon_{n,k_y}  \hat{c}_{n\sigma,k_y}^{\dagger} \hat{c}_{n\sigma',k_y} \nonumber \\
\mbox{} && - \gamma \sum_{n,\sigma, k_y,\alpha=x,z}  (\hat{c}_{n\sigma,k_y}^{\dagger}\hat{c}_{n + \mathbf{e}_\alpha,\sigma,k_y} + \mathrm{H.c.}).
\end{eqnarray}
where the operators $\hat{c}_{n \sigma}^\dagger$ ($\hat{c}_{n \sigma}$) create (annihilate) electron with spin $\sigma$ on site $n$ with the transverse momentum $k_y$. The kinetic energy $\varepsilon_{n,k_y} = - 2 \gamma \cos k_y a$ is equivalent to an increase in the on-site energy, and the nearest neighbor hopping is set at \mbox{$\gamma=1.0$ eV}.

The evaluation of Eq.~\eqref{eq:lbsee} of relies crucially on the construction of the proper coupling matrix ${\bm \tau}$ between $\hat{H}_\mathrm{TI}$ in Eq.~\eqref{eq:ti} and $\hat{H}_N$ in Eq.~\eqref{eq:n} since ${\bm \tau}$  enters into the retarded GF. For example, the Hamiltonian of the composite system semi-infinite-N-lead-1 + TI-thin-film is given by
\begin{equation}\label{eq:composite}
\mathbf{H}_{N+\mathrm{TI}} = \left( \begin{array}{cc}
\mathbf{H}_N & {\bm \tau}_1 \\
{\bm \tau}^{\dagger}_1 & \mathbf{H}_\mathrm{TI}
\end{array} \right).
\end{equation}
The retarded GF of the TI film alone, viewed as an \emph{open} quantum system, is defined by~\cite{Datta1995,Stefanucci2013}
\begin{equation}\label{eq:retardedgf}
\mathbf{G}(E)=[E-\mathbf{H}_\mathrm{TI}-{\bm \Sigma}_1(E)]^{-1}.
\end{equation}
Here the retarded self-energy introduced by the semi-infinite N lead 1 is
\begin{equation}
{\bm \Sigma}_1(E)={\bm \tau}_1^\dagger \cdot \mathbf{g}(E) \cdot {\bm \tau}_1,
\end{equation}
and ($\eta$ is positive infinitesimal)
\begin{equation}
\mathbf{g}(E) = [E + i\eta - \mathbf{H}_{N}]^{-1},
\end{equation}
is the retarded GF~\cite{Datta1995,Stefanucci2013} of N lead 1. The same procedure would be repeated when more than one N lead is attached to
the TI thin film to get
\begin{equation}
\mathbf{G}(E)=[E-\mathbf{H}_\mathrm{TI}-{\bm \Sigma}_1(E) - {\bm \Sigma}_2(E) - {\bm \Sigma}_3(E)]^{-1},
\end{equation}
for the three-terminal junction in Fig.~\ref{fig:fig1}.

The conventionally assumed identical orientation of spin (i.e., expectation value of the spin operator) on Bi and Se sublattices, where spin on both sublattices follows ``helical'' texture shown in Fig.~3 for Bi sublattice, is valid only on the $(111)$ surface of the TI crystal that coincides with the plane of the QL. For TI surface other than (111), the spin operators for electrons residing on the Bi and Se sublattices are inequivalent.~\cite{Silvestrov2012} Nevertheless, for interpreting spin- and angular- resolved photoemission spectroscopy (spin-ARPES) experiments~\cite{Hasan2010}  or for attaching the TI sample to N leads, where photoelectrons or electrons injected or absorbed from N leads do not carry a sublattice index, it is advantageous to select the standard relation between the Pauli matrices and the spin operator, $\hat{\bf S} = \hbar \hat{\bm \sigma}/2$.

\begin{figure}[b]
\includegraphics[scale=0.6,angle=0]{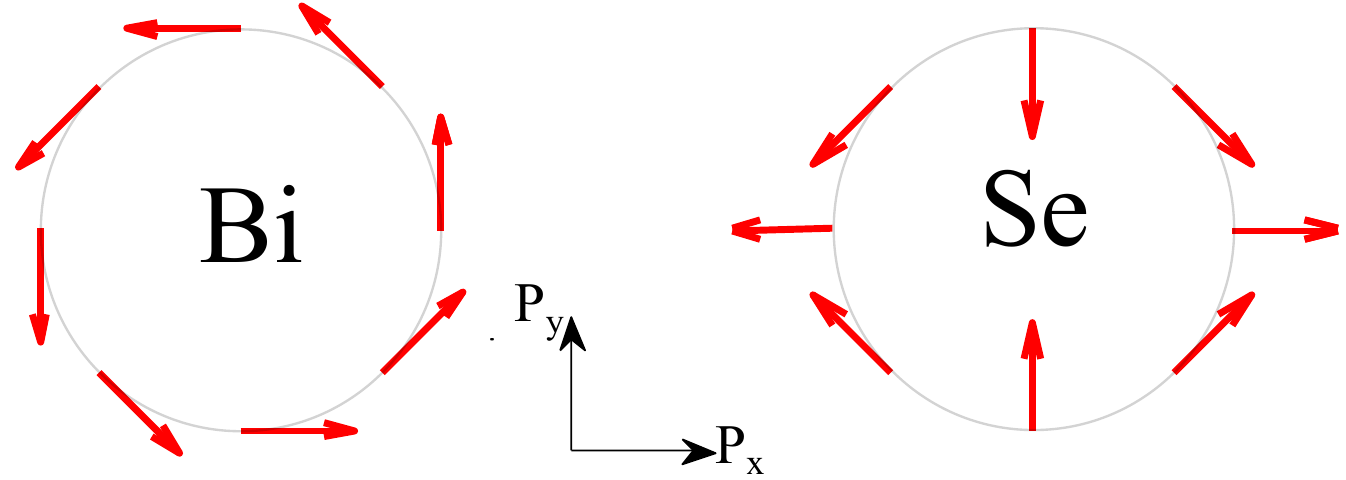}
\caption{(Color online) The equilibrium expectation values of the spin operator for surface-state electrons on the Bi and Se sublattices as a function of the in-plane momentum \mbox{$\mathbf{p}=(p_x,p_y)$}. The Bi$_2$Se$_3$ crystal is assumed to fill the half-space $z<0$, so that its infinite surface in the $xy$-plane is {\em orthogonal} to its QLs that are also oriented perpendicularly to the unit vector $\mathbf{n}=(\mathbf{e}_x + \mathbf{e}_y)/\sqrt{2}$, as illustrated in  Fig.~\ref{fig:fig1}.}
\label{fig:fig3}
\end{figure}
\begin{figure*}
\includegraphics[scale=0.5,angle=0]{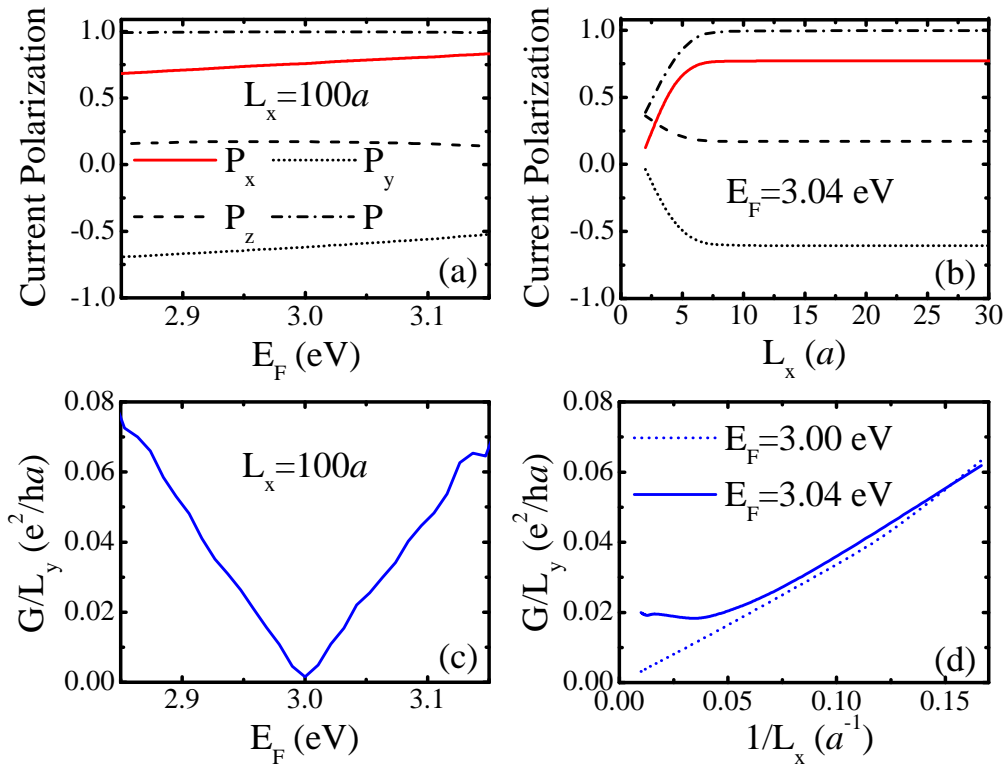}
\caption{(Color online) The components $(P_x,P_y,P_z)$ and the magnitude $|\mathbf{P}|$ of the spin-polarization vector of charge current outflowing into lead 2, after unpolarized charge current is injected from lead 1 into the two-terminal version of junction in Fig.~\ref{fig:fig1}. These are plotted vs. (a) the Fermi energy $E_F$,  or (b) the length $L_x$ of the TI film. The linear response charge conductance of the same two-terminal junction vs. (c) the Fermi energy, or (d) the inverse length $1/L_x$.}
\label{fig:fig2}
\end{figure*}

The spinors $u_\mathrm{Bi}$ and $u_\mathrm{Se}$, associated with each sublattice when inequivalent spin operators are used, have to be unitarily transformed~\cite{Silvestrov2012} to \mbox{$u_\mathrm{Bi} \mapsto u_\mathrm{Bi}$} and \mbox{$u_\mathrm{Se} \mapsto i (\hat{\bm \sigma} \cdot \mathbf{n}) u_\mathrm{Se}$}. Here $\mathbf{n}$ is the unit vector normal to the QL. This specifies  $2 \times 4$ coupling matrices for each N lead \mbox{$p=$1--3} as
\begin{equation}\label{eq:tau}
{\bm \tau}_p=\left( \begin{array}{cccc}
t_\mathrm{Bi} & 0 & i n_zt_\mathrm{Se} & (in_x+n_y) t_\mathrm{Se} \\
0 & t_\mathrm{Bi} & (in_x - n_y) t_\mathrm{Se} & -i n_z t_\mathrm{Se}
\end{array} \right).
\end{equation}
For the setup in Figs.~\ref{fig:fig1} and ~\ref{fig:fig3} we use $\mathbf{n}=(110)$ in Eq.~\eqref{eq:tau}. In addition, the hopping parameters between orbitals in the N leads and those on the Bi or Se sublattice are chosen as: \mbox{$t_\mathrm{Bi} = 0.4$} eV for leads 1 and 2; \mbox{$t_\mathrm{Bi} = 0.45$ eV} for lead 3; \mbox{$t_\mathrm{Se} = 0.8$ eV} for leads 1 and 2; and  \mbox{$t_\mathrm{Se} = 0.9$ eV} for lead 3.

\section{Nonequilibrium spin polarization in two-terminal TI-based junctions}\label{sec:polarization}

When unpolarized charge current is injected from N lead 1, the ensemble of outgoing spins in N lead 2 of N$_1$/TI/N$_2$ two-terminal junction  is characterized by the spin density matrix \mbox{$\hat{\rho}^{\rm out}_{\rm spin} = \frac{1}{2}(1+{\bf P} \cdot \hat{\bm \sigma})$} whose polarization vector is given by $P_\alpha = I_p^{S_\alpha}/I_p$ for such setup. We first demonstrate in Fig.~\ref{fig:fig2}(a) that an unpolarized charge current injected from N lead 1 into the top surface of a two-terminal junction (i.e., when the third N lead in Fig.~\ref{fig:fig1} is removed) will exit into N lead 2 with non-zero spin-polarization vector $\mathbf{P}=(P_x,P_y,P_z)$ which includes a  component $P_x \neq 0$ in the direction of transport. Since Dirac fermions on the opposite surfaces of TI have opposite chiralities, which generates opposite spin-polarization for currents flowing through the top and bottom surface that would cancel in the total current in N lead 2, we block transport through the bottom surface by introducing an energy gap into its Dirac cone (e.g., due to coating by magnetic film~\cite{Luo2013}). Figure~\ref{fig:fig2}(b) reveals that current spin-polarization is established on a very short length scale of $\simeq 10$ lattice spacings, so that this mechanism can operate near or under the contacts with N leads even in the presence of inevitable spin or charge dephasing mechanisms (the spin dephasing time for the in-plane spin components on the surface of TI is the same as the momentum relaxation time~\cite{Burkov2010}).

Although the $g$-factor in the Zeeman term $-g_\parallel \mu_B B_x \hat{\sigma}_x$ introduced by the external magnetic field applied parallel to the top and bottom surfaces of Bi$_2$Se$_3$ is renormalized~\cite{Winkler2003} $g_\parallel=23$ due to strong SOC effects,~\cite{Kohler1975} changing its sign \mbox{$B_x \rightarrow -B_x$} has virtually no effect on the spin-polarization vector governed by the strong surface SOC. For example, \mbox{$P_x=0.605$} (at $E_F=3.04$ eV selected for illustration) in zero magnetic field \mbox{$B_x =0$}  is virtually indistinguishable from \mbox{$P_x=0.609$} at large external magnetic field \mbox{$B_x = -10$} T applied opposite to the direction of electron transport. Thus, the same mechanism---momentum-dependent effective magnetic field associated with SOC which is much stronger than any external one---can be invoked to explain why $V_\mathrm{ISHE}$ did not change sign upon reversing $B_x \rightarrow -B_x$ in the experiment of Ref.~\onlinecite{Jaworski2012}. This requires that 2D hole gas (see Fig. 1 in Ref.~\onlinecite{Jaworski2012}) formed at the interface between Pt probe and InSb has SOC with a component of its effective  magnetic field pointing in the direction of transport (as it would be the case in the presence of the Dresselhaus SOC~\cite{Winkler2003}).

At first sight, the surface of 3D TI is expected to spin-polarize charge current in the transverse direction only,~\cite{Burkov2010,Misawa2011,Pesin2012,Modak2012} $P_y \neq 0$ while $P_x=P_z=0$. This is due to the fact that Dirac cone energy-momentum dispersion on the surface of TI and spin-orthogonal-to-momentum locking within it, as observed in spin-ARPES experiments,~\cite{Hasan2010} is routinely described by an effective 2D Hamiltonian~\cite{Hasan2010,Liu2010} taking form of the massless Rashba model,~\cite{Winkler2003} $\hat{H} = v_F (\hat{\bm \sigma} \times \hat{\mathbf{p}}) \cdot \mathbf{e}_z$. Here $v_F$ is the Fermi velocity, $\hat{\mathbf{p}}=(\hat{p}_x, \hat{p}_y)$ is the momentum operator in 2D and $\hat{\bm \sigma} = (\hat{\sigma}_x,\hat{\sigma}_y,\hat{\sigma}_z)$ is the vector of the Pauli matrices. Thus, when charge current flows on the TI surface longitudinally, this Hamiltonian predicts induction of {\em nonequilibrium} transverse spin density~\cite{Burkov2010,Misawa2011,Pesin2012} $S_y$ and the corresponding spin-polarization $P_y \neq 0$ of the current outflowing into the attached N leads.~\cite{Modak2012} This effect is the counterpart of the one predicted long ago,~\cite{Edelstein1990,Inoue2003} and observed in recent experiments,~\cite{Ganichev2006a} for the Rashba spin-split 2DEGs, except that $S_y$ on the TI surface is larger by a factor $\hbar v_F/\alpha \gg 1$ (the Rashba SOC term in 2DEGs is given by~\cite{Winkler2003} $\alpha (\hat{\bm \sigma} \times \hat{\mathbf{p}}) \cdot \mathbf{e}_z/\hbar$).

\begin{figure*}
\includegraphics[scale=0.5,angle=0]{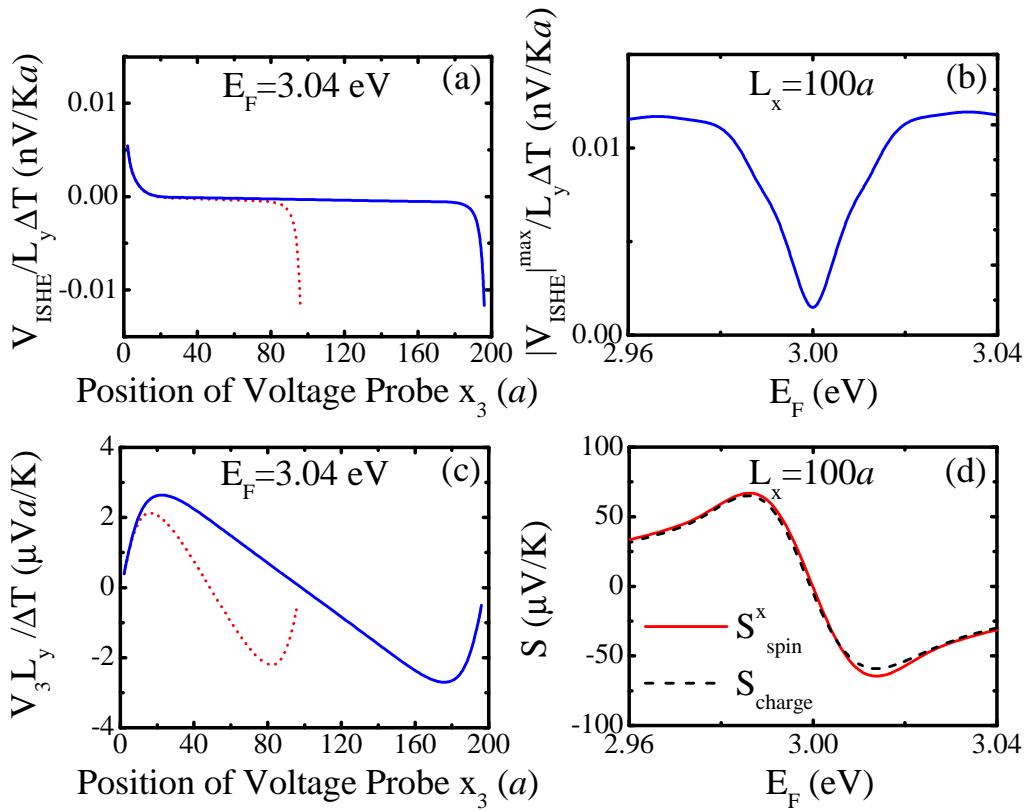}
\caption{(Color online) (a) The SSE signal $V_\mathrm{ISHE}/(L_y \Delta T)$ as a function of the position of N lead 3 in Fig.~\ref{fig:fig1} displaced between the contacts with N leads 1 and 2 sandwiching TI samples of length $L_x=100a$ or $L_x=200a$. (b) Dependence of $|V_\mathrm{ISHE}|^\mathrm{max}/(L_y \Delta T)$ on the Fermi energy $E_F$. (c) Voltage $V_3L_y/\Delta T$ applied to N lead 3 in order to bring the net charge current $I_3 \equiv 0$. (d) Conventional charge and spin-dependent Seebeck coefficients vs. $E_F$ for the two-terminal version of junction in Fig.~\ref{fig:fig1}.}
\label{fig:fig4}
\end{figure*}

However, such conclusion is an artifact of a na\"{i}ve identification of $\hat{\bm \sigma}$ operator with the true electron spin, which becomes invalid
when the TI surface does not coincide with the QL plane.~\cite{Silvestrov2012} Since low-energy Hamiltonian models~\cite{Liu2010} of Bi$_2$Se$_3$ operate with the pseudospin degree of freedom describing states with support on the Bi and Se sublattices, the attachment of the TI thin film to semi-infinite N leads that inject or absorb electrons carrying real spin requires to properly interpret their mutual coupling when studying spin-dependent electron transport. Aligning QLs of Bi$_2$Se$_3$ perpendicularly to the infinite TI surface in $xy$-plane, and at an angle of $45^\circ$ with respect to the $yz$-plane, will generate difference in spin textures on the two sublattices shown in Fig.~\ref{fig:fig3}. This motivates our proposal for the junction setup in Fig.~\ref{fig:fig1}, where longitudinal spin polarization [$P_x \neq 0$ in Fig.~\ref{fig:fig2}(a) and (b)] is driven by the surface-state electrons on the Se sublattice. This emerges in addition to the amply studied (in the diffusive~\cite{Burkov2010,Misawa2011,Pesin2012} or in the ballistic~\cite{Modak2012} transport regimes) transverse nonequilibrium spin density and polarization [$P_y \neq 0$ in Fig.~\ref{fig:fig2}(a) and (b)] that is predominantly generated by the Bi sublattice. Note that  $P_z$ component in Fig.~\ref{fig:fig2}(a) and (b) remains non-zero even if both surfaces are open for transport, or if the cross section in the $yz$-plane becomes infinite, since it  originates from electrons tunneling through the bulk of the TI.

Figure~\ref{fig:fig2}(c) plots the linear-response conductance \mbox{$G = \lim_{V_b \rightarrow 0} I_2/V_b$} of the TI thin film attached to two N leads as a function of the Fermi energy $E_F$ when small bias voltage $V_b=V_1-V_2$ drives charge current $I_2$. The $G$ vs. $E_F$ dependence exhibits a V-shape (slightly  asymmetric due to the attached N leads) familiar from graphene,~\cite{Zuev2009,Tworzydlo2006} with a minimum conductivity $\sigma = G L_x/L_y$ reached at the DP located at \mbox{$E_F = 3.0$ eV}. Even though the density of states vanishes at the DP, so that $\sigma$ should apparently approach zero at the DP, for this {\em ballistic} junction it remains non-zero due to evanescent  wavefunctions injected by the metallic N leads. While they are similar to the well-known metal induced gap states in metal-semiconductor junctions, such states typically penetrate only a few atomic lengths into the semiconductor where the depth of penetration decreases with increasing band gap. On the other hand, evanescent states in N/TI junctions penetrate a much longer distance due to zero energy gap at the DP, as observed also in N/graphene junctions.~\cite{Golizadeh-Mojarad2009} Figure~\ref{fig:fig2}(d) shows accidental ($\beta \approx 1$) Ohmic scaling $G \propto L_y/L_x^\beta$ (for $L_y/L_x \gg 1$) at DP, so that evanescent mode quantum transport in N/TI or N/graphene junctions is termed~\cite{Tworzydlo2006}  ``pseudo-diffusive.''

\section{Voltage signal of SSE in three-terminal TI-based junctions}\label{sec:sse}

When the third N lead, assumed to be made of a heavy metal with sufficiently large~\cite{Liu2012} SH angle $\theta_\mathrm{SH}$, is attached to the top surface of the TI, as shown in Fig.~\ref{fig:fig1}, spin current $I_3^{S_\alpha}$ will be injected into it. Besides using temperature bias \mbox{$\Delta T = T_1 - T_2= 2$ K} at average temperature \mbox{$T=(T_1+T_2)/2=50$ K} to drive SSE, we also apply voltage $V_3$ to the macroscopic reservoir (attached to N lead 3 at infinity) in order to ensure that net charge current through it remains zero $I_3 \equiv 0$ and $I_3^{S_\alpha} \neq 0$ is pure. The profile of $V_3$ across the TI thin film is plotted in Fig.~\ref{fig:fig4}(c). We assume that the reservoir temperature $T_3 (x)=T_1 - x(T_1-T_2)/L_x$ decreases linearly as N lead $3$ is displaced between the contacts of TI film with N lead 1 and 2.

The central quantity in the theories~\cite{Adachi2013} of transverse SSE is $I_3^{S_x}$ component of the pure spin current injected into N lead 3, which we compute per lattice spacing $a$ because of assumed periodicity of system in Fig.~\ref{fig:fig1} in the $y$-direction. Since $I_3^{S_x}$ carries spins pointing along the $x$-axis, the ISHE mechanism illustrated in Fig.~\ref{fig:fig1} will accumulate charges on the opposite edges of N lead 3 in the transverse direction. These generate electric field $\mathbf{E}_\mathrm{ISHE}$ and the corresponding voltage signal~\cite{Adachi2013} \mbox{$V_\mathrm{ISHE} = E_\mathrm{ISHE}^y L_y = \theta_\mathrm{SH}^\mathrm{Pt} I_3^{S_x} e \rho^\mathrm{Pt}/W$}. To facilitate comparison with experiments,~\cite{Jaworski2010,Jaworski2012} Fig.~\ref{fig:fig4} plots $V_\mathrm{ISHE}/(L_y\Delta T)$ which has the same unit (after multiplying the results in Fig. 4 by $L_y$) as the conventional charge-Seebeck coefficient $S_\mathrm{charge} = - (V_1 - V_2)/(T_1 - T_2)$ measured~\cite{Zuev2009} on two-terminal junctions. For this purpose, we assume that $I_3^{S_x}$ is converted into $V_\mathrm{ISHE}$ via the ISHE operating within N lead 3 of width $W=6a$ which is made of Pt with resistivity $\rho^\mathrm{Pt} = 105$ n$\Omega$m and with putative~\cite{Liu2012} SH angle   $\theta_\mathrm{SH}^\mathrm{Pt}=0.08$.

The dependence of $V_\mathrm{ISHE}/(L_y \Delta T)$ on the position of N lead 3 shown in Fig.~\ref{fig:fig4}(a) is asymmetric, thereby exhibiting a {\em fundamental property} of the transverse SSE where its signal changes sign between cold and hot ends of the sample.~\cite{Uchida2012,Jaworski2010,Jaworski2012} The maximum signal $|V_\mathrm{ISHE}|^\mathrm{max}/(L_y \Delta T)$ is reached around the sample edges and it is \emph{independent} of length $L_x$, as shown in Fig.~\ref{fig:fig4}(a). Unlike the spin-dependent Seebeck coefficient $S^x_\mathrm{spin}$ plotted in Fig.~\ref{fig:fig4}(d), which quantifies spin current injected into N lead 2 due to temperature bias applied to the two-terminal version of junction in Fig.~\ref{fig:fig1},  $|V_\mathrm{ISHE}|^\mathrm{max}/(L_y \Delta T)$ vs. $E_F$ within the bulk gap of TI plotted in Fig.~\ref{fig:fig4}(b) is unrelated to conventional CS coefficient $S_\mathrm{charge}$. Note that both $S_\mathrm{spin}^x$ and $S_\mathrm{charge}$ in Fig.~\ref{fig:fig4}(d) exhibit the same dependence on $E_F$ as $S_\mathrm{charge}$ measured on graphene.~\cite{Zuev2009} This is due to the fact that electron- and hole-like transport gives contributions to these coefficients of opposite sign, so that \mbox{$S_\mathrm{spin}^x = S_\mathrm{charge} \equiv 0$} exactly at the DP while reaching maximum absolute value few $k_BT$ away from it.

\section{Concluding remarks}\label{sec:conclusions}

In conclusion, we predict that thermally driven charge current on the surface of 3D TI thin film, realized using Bi$_2$Se$_3$ whose QLs are oriented at an angle of $45^\circ$ with respect to the direction of transport while being perpendicular to the TI surface (see Fig.~\ref{fig:fig1}), will become spin-polarized due to \emph{strong surface SOC}. In addition to amply studied~\cite{Burkov2010,Misawa2011,Pesin2012,Modak2012} nonequilibrium transverse spin polarization on the TI surface parallel to QLs, for the orientation of QLs we propose in Fig.~\ref{fig:fig1} the spin-polarization vector will acquire an additional component parallel to the direction of charge transport which is generated by the Se sublattice. This makes it possible to recreate the phenomenology of {\em electron-driven} SSE recently observed in InSb,~\cite{Jaworski2012} but in the \emph{absence} of any external magnetic field.  The predicted magnitude of SSE signal shown in Figs.~\ref{fig:fig4}(a) and ~\ref{fig:fig4}(b) can be translated into experimentally measurable voltage by multiplying it with $L_y$ and $\Delta T$, e.g., \mbox{$|V_\mathrm{ISHE}|^\mathrm{max} \simeq 5$ $\mu$V} if we assume \mbox{$L_y = 1$ mm}, \mbox{$\Delta T=2$ K} and \mbox{$a \approx 3.4$ nm} (which is the effective lattice constant of our simple cubic lattice for the distance \mbox{$c \approx 2.9$ nm} between QLs oriented as in Fig.~\ref{fig:fig1}) . We believe that this value could be further enhanced by including phonon-electron drag effect, speculated to play a key role in achieving the ``giant'' magnitude of SSE measured in Ref.~\onlinecite{Jaworski2012}, which we relegate to future studies.

\begin{acknowledgments}
We thank E. Saitoh for illuminating discussions. P.-H. C., F. M. and B. K. N. were supported by NSF under Grant No. ECCS 1202069. N. N. was
supported by Grant-in-Aids for Scientific Research (21244053) from the Ministry of Education, Culture, Sports, Science and
Technology of Japan, Strategic International Cooperative Program (Joint Research Type) from Japan Science and
Technology Agency, and also by Funding Program for World-Leading Innovative R\&D on Science and Technology (FIRST Program).
\end{acknowledgments}




\end{document}